\documentclass[aps,pre,twocolumn,superscriptaddress,showpacs]{revtex4}

\usepackage[dvips]{graphicx}
\usepackage{amsmath}
\usepackage{amssymb}
\usepackage{bm}

\begin{document}
\title{Nature of perturbation theory in
spin glasses}

\author{J.  Yeo}  \affiliation{School of Physics  and  Astronomy,
University    of   Manchester,   Manchester    M13   9PL,    U.\   K.}
\affiliation{Department of Physics,  Konkuk University, Seoul 143-701,
Korea}
\author{M. A. Moore} \affiliation{School of Physics and Astronomy,
University of Manchester, Manchester M13 9PL, U.\ K.}
\author{T. Aspelmeier} \affiliation{Institut f\"ur Theoretische
Physik,  Georg-August-Universit\"at,  D37077, G\"ottingen,  Germany }
\date{\today}

\begin{abstract}
The  high-order behavior of  the perturbation  expansion in  the cubic
replica field  theory of  spin glasses in  the paramagnetic  phase has
been investigated.  The study starts with the zero-dimensional version
of the replica field theory and  this is shown to be equivalent to the
problem  of finding finite  size corrections  in a  modified spherical
spin glass near the critical  temperature. We find that the high-order
behavior of the perturbation series is described, to leading order, by
coefficients  of alternating  signs (suggesting  that the  cubic field
theory is well-defined) but that there are also subdominant terms with
a complicated dependence of their  sign on the order.  Our results are
then extended  to the $d$-dimensional  field theory and  in particular
used  to  determine  the  high-order  behavior of  the  terms  in  the
expansion   of  the   critical  exponents   in  a   power   series  in
$\epsilon=6-d$.   We  have  also  corrected  errors  in  the  existing
$\epsilon$ expansions at third order.

\end{abstract}


\maketitle


\section{introduction}

The  theory  of  spin   glasses  in  finite  dimensional  systems  has
traditionally been  approached by  the loop expansion  around Parisi's
mean-field replica  symmetry breaking solution  \cite{P}. However, the
picture of spin glasses  which emerges from this perturbative approach
is   quite  different   to   the  one   arising   in  droplet   theory
\cite{droplet}.   The   motivation  for  the  present   paper  was  to
investigate the  possibility that perturbation theory  in spin glasses
might fail  for some reason (for example,  non-perturbative terms like
``droplets''  might  dominate  their  free  energy, at  least  in  the
low-temperature phase).  We have started the programme with a study of
the nature  of the perturbation  expansion in the  high-temperature or
paramagnetic phase,  and postpone to  another paper the  discussion of
the low-temperature phase.

In general the nature  of perturbation expansions in disordered system
is far from  trivial.  This is in contrast  to the perturbation theory
in  pure systems,  where, for  example, the  Borel summability  of the
series  leads to  an  accurate evaluation  of  critical  exponents
\cite{LZ}.  In the study  of the perturbation expansion for disordered
ferromagnets,  Bray  et al.\  \cite{BMMRY}  found  that  even in  zero
dimensions, the high-order behavior  of the perturbation expansion was
surprisingly rich.  The high-order  expansion coefficients are sums of
two kinds  of terms: one  type has an unusual  cosine-like oscillation
with  increasing  periodicity  and   the  second  type  has  a  simple
alternation in  sign which dominates for small  disorder. This unusual
behavior        has        been        further       studied        in
Refs.~\cite{McKane94},~\cite{Alvarez}, with  the final conclusion that
the  series is  still summable,  but that  the simple  Borel procedure
needs to be modified to  deal successfully with long series.  We shall
find that  the perturbation expansion for spin  glasses has remarkably
similar features to those of the disordered ferromagnet.

Our investigation of the nature  of the perturbation expansions in the
high-temperature  phase  of spin  glasses  starts  by  looking at  the
perturbation  expansion  of the  zero-dimensional  spin glass  problem
(which   will  be   referred  to   as  the   ``toy''   problem).   The
zero-dimensional  field theory  is  the  key to  the  analysis of  the
$d$-dimensional field  theory as the extension  to the $d$-dimensional
field theory  and critical  exponents is a  relatively straightforward
extension of  the toy problem \cite{McKane}.  Apart  from being simple
integrals, the zero-dimensional toy  field theory has the advantage of
allowing an analysis without replicas.  In this paper this is achieved
by mapping the problem to that of critical finite-size scaling in a
modified  version of  the  spherical spin  glass  \cite{KTJ} and  this
mapping allowed us to find  the high-order behavior without the use of
replicas.   However,   to  obtain  the  high-order   behavior  of  the
perturbation expansion  for the $d$-dimensional  field theory requires
the  use of  replicas  and we  found that  this  needed the  use of  a
non-trivial replica  symmetry breaking scheme in the toy model in order  to get results
consistent with our mapping to  the spherical model.  Our chief result
is  that the  perturbation  theory is  well-defined  and the  dominant
high-order terms  in the  perturbation expansion have  coefficients of
alternating   signs.   However,  the   perturbation   series  of   the
zero-dimensional spin  glass field theory  is not Borel summable  in a
straightforward way due to the presence of subdominant terms.

The paper is organized as follows. In the next section, we consider
the cubic replica field theory of spin glasses and obtain the first few
expansion coefficients of the zero-dimensional
toy problem by explicitly evaluating the Feynman diagrams. In Sec.~\ref{sp},
we show the equivalence of the zero-dimensional field theory and
the critical finite-size scaling of the modified spherical model.
Using this mapping we obtain the high order behavior of the
perturbation expansion for the toy problem in Sec.~\ref{analysis}.
In Sec.~\ref{ToyReplica}, we consider the toy problem using  replicas.
This is generalized to the problem of the high order terms
in the $\epsilon$ expansion in Sec.~\ref{epsilon}.
We conclude with a discussion in Sec.~\ref{conclusion}.

\section{Replica field theory of spin glasses}
\label{zero}

The replica  field theory of spin  glasses, (see Ref. \cite  {KD} for a
review), starts from the Hamiltonian density
\begin{align}
&\mathcal{H}=\frac 14\sum_{\alpha,\beta} (\nabla q_{\alpha\beta})^2
           +\frac \tau 4\sum_{\alpha,\beta}q_{\alpha\beta}^2
           -\frac w6
\sum_{\alpha,\beta,\gamma}q_{\alpha\beta}q_{\beta\gamma}q_{\gamma\alpha}
\label{FT}\\
&-y\Big(\frac 1 {12} \sum_{\alpha,\beta}q_{\alpha\beta}^4
   +\frac 1 8
    \sum_{\alpha,\beta,\gamma,\delta}q_{\alpha\beta}q_{\beta\gamma}
    q_{\gamma\delta}q_{\delta\alpha}-\frac 1 4 \sum_{\alpha,\beta,\gamma}
               q_{\alpha\beta}^2 q_{\alpha\gamma}^2 \Big).\nonumber
\end{align}
As usual the field components $q_{\alpha\beta}$ ($\alpha\neq\beta$ and 
$q_{\alpha\beta}=q_{\beta\alpha}$)
take all real values, and
the indices such as $\alpha$ take the values $1, 2, 3, \hdots, n$. In the limit
when $n$ goes to zero, such a field theory is thought to capture the physics of
finite dimensional spin glasses.
The quartic terms work as stabilizing terms, but for $d<6$ are irrelevant
variables which we shall drop. One question which we shall ask is whether
the resulting cubic field theory is well-defined in the limit
$n\rightarrow 0$.
It is possible this approach  may not be valid
as  cubic  theories have  Hamiltonians  which  are  not bounded  below
\cite{McKane}.  However, there are examples of field theories existing
where unphysical limits, such as the number of field components $n$ is
taken to zero, which  saves these apparently unphysical theories \cite{McKane, BM}. 
This seems also to be the case for spin glasses as our work shows that the
coefficients in the perturbation expansion in $w$ alternate in sign, which is an indication
that the field theory remains well-defined in the limit when $n$ goes to zero.

The partition function of the zero-dimensional spin glass
field theory is given by
\begin{eqnarray}
Z&=&\int \prod_{\alpha<\beta} \left(\frac{dq_{\alpha\beta}}
{\sqrt{2\pi}}\right)\exp\Big[
-\frac \tau 4 \sum_{\alpha,\beta}q_{\alpha\beta}^2 \nonumber \\
&&\quad\quad\quad +\frac w6
\sum_{\alpha,\beta,\gamma}q_{\alpha\beta}q_{\beta\gamma}q_{\gamma\alpha}
\Big].
\label{Zn}
\end{eqnarray}
The perturbation expansion is well-defined irrespective of whether the
integral of Eq.~(\ref{Zn}) actually exists.
The perturbation expansion in $w$ yields a series
\begin{equation}
Z(g^2)=\tau^{-\frac{n(n-1)}4}\left[1+\sum_{K=1}^\infty A_K
g^{2K} \right],
\label{seriesZn}
\end{equation}
where we take $g^2=w^2/ (\tau/2)^3$ as the expansion parameter of the problem.
The use of $\tau/2$ instead of $\tau$ is for later convenience.
The series expansion of the corresponding
free energy is given by
\begin{equation}
\beta F(g^2)=\frac{n(n-1)}4 \ln\tau-
\sum_{K=1}^\infty B_K g^{2K}.
\end{equation}

\begin{figure}
\includegraphics[width=0.4\textwidth]{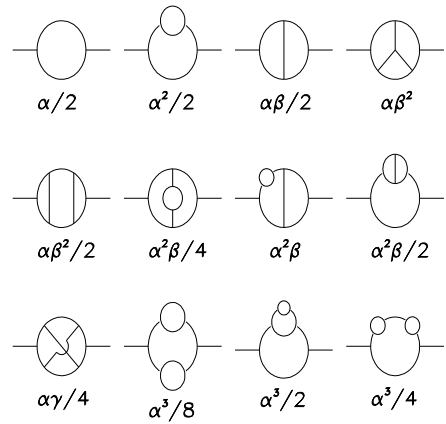}
\caption{Feynman diagrams to O($g^8$) contributing to the renormalized propagator
with their corresponding contributions.}
\label{prop}
\end{figure}

\begin{figure}
\includegraphics[width=0.5\textwidth]{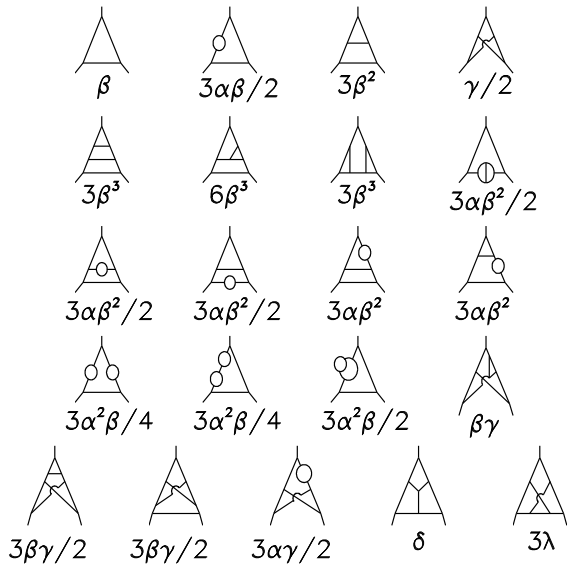}
\caption{Feynman diagrams to O($g^8$) contributing to the renormalized vertex
with their corresponding contributions.}
\label{vertex}
\end{figure}

Although this zero-dimensional theory is
nothing but a multiple integral, it is not an easy task
to calculate the expansion
coefficients $A_K$ or $B_K$ directly by expanding
the exponential in (\ref{Zn}),
because of the complicated structure of the internal summations.
In fact, even using a symbolic
manipulation program on a computer, we find it
very difficult to get the expansion coefficients
higher order than the first couple of terms.
Fortunately, in the study of a 
cubic field theory similar to ours
-- the percolation problem \cite{AKM} --
different types of the internal contractions occurring in the theory
were classified diagrammatically,
and all the relevant diagrams were given up to O($g^8$).
To this order there are five different internal contraction types.
(See the diagrams denoted by $\alpha$, $\beta$,
$\gamma$, $\delta$ and $\lambda$ in Figs.~\ref{prop} and \ref{vertex}.)
These were translated into the spin glass problem in Ref.~\cite{Green}.
In Figs.~\ref{prop} and \ref{vertex}, all the diagrams to this
order contributing to the renormalized propagator
and vertex are listed along with their contributions. Note that,
in the zero-dimensional field theory, each contribution is just given
by the product of these contraction factors.

The expansion coefficient $B_K$ of the free energy can be obtained by
noting that $-\partial (\beta F)/\partial w$ is just the renormalized
three-point function, which can be written in terms of the renormalized propagator
and vertex.
Collecting the contribution from each diagram
and using the results of the internal contractions, we
calculate the perturbation expansion coefficient $B_K$ up to
four-loop order (0$(g^8)$). We find that
$B_K=\frac 1 6 n(n-1)(n-2)f_K/8^{K}$, where
$f_1=1/2$, $f_2=n-2$, $f_3=4n^2-31n/2+44/3$
and $f_4 =22n^3-123n^2+229n-148$.
Therefore, the free energy for the toy spin glass
problem, $\lim_{n\to 0} F/n$, is given by
a power series in $g^2$ with coefficients of alternating signs up to O$(g^8)$.

In the course of investigation,
we have discovered errors in two ($\delta$ and $\lambda$
in Figs.~\ref{prop} and \ref{vertex})
of the five types of contraction
reported in Ref.~\cite{Green}.
The correct results we obtain are $\delta=n^3-9n^2+54n-104$
and $\lambda=5n^2-14n$, while the results in Ref.~\cite{Green}
were $\delta=n^3-3n^2+38n-94$ and $\lambda=5n^2-2n-12$. The latter
can easily be shown to be incompatible with the simple $n=3$ case.
These corrections will change the O($\epsilon^3$) terms
in the critical exponents in $d=6-\epsilon$ dimensions.
From the explicit expressions for the critical exponents $\eta$
and $\nu$ obtained in Ref.~\cite{AKM}, we calculate the correct form
of the $\epsilon$ expansion to third order to be
\begin{eqnarray}
&&\eta=-0.3333\epsilon+1.2593\epsilon^2+0.7637\epsilon^3  \label{eta}\\
&&\nu^{-1}-2+\eta=-2\epsilon+9.2778\epsilon^2-6.4044\epsilon^3. \label{nu}
\end{eqnarray}
The corrected series for $\nu^{-1}-2+\eta$ shows an oscillation in signs
in contrast to the one in Ref.~\cite{Green} where the $O(\epsilon^3)$ term
was positive.

\section{mapping of the toy problem onto a modified spherical model}
\label{sp}

As mentioned in the  Introduction, we study the zero-dimensional field
theory by  mapping it  onto a modified  version of the  spherical spin
glass model. The Hamiltonian of the spherical model is
\begin{equation}
H_{\mathrm{sp}}=-\frac 1 2 \sum_{i,j} J_{ij} S_i S_j, 
\end{equation}
with the spherical constraint $\sum_i S^2_i =N$ among the $N$ spins.
Unlike the conventional spherical spin
glass model \cite{KTJ}, we take $J_{ii}\neq 0$ in addition to the 
infinite-ranged interactions
$J_{ij}=J_{ji}$ ($i\neq j$). They are chosen
from Gaussian distributions
\begin{equation}
 P(J_{ij}) \sim e^{-\frac N 4 \mathrm{tr} \mathbf{J}^2}
             =\prod_i e^{-\frac N 4 J^2_{ii}}\prod_{i<j} e^{-\frac N 2 J^2_{ij}}.
\end{equation}
The presence of the diagonal interaction does not make any difference in the $N\to\infty$
limit. In the following, however, we consider finite size corrections in this model.
The partition function can be written as
\begin{eqnarray}
Z_{\mathrm{sp}}
&=&\frac\beta 2 \int_{-\infty}^{\infty}\prod_i dS_i \int_{-i\infty}^{i\infty}
   \frac{dz}{2\pi i} \nonumber \\
&\times & \exp\Big[ \frac \beta 2
          \big( Nz-z\sum_i S^2_i +\sum_{i,j} J_{ij}S_i S_j \big) \Big],
\label{Zsph}
\end{eqnarray}
where $\beta=1/T$ is the inverse temperature and
the chemical potential $z$ was introduced to represent the delta function
$\delta (N-\sum_i S^2_i)$.

In order to make a connection to the replica field theory (\ref{Zn}),
we replicate the partition function $n$ times and average over the Gaussian bond
distribution. We then take the usual Hubbard-Stratonovich (HS) transformations
on the factor 
$\exp[(\beta^2/4N)\sum_{\alpha,\gamma}(\sum_i S^\alpha_i
S^\gamma_i)^2]$
to get the spins on the same site, where the Greek indices denote the replica
components. In order to do that, we need to introduce 
the diagonal $q_{\alpha\alpha}$ and off-diagonal $q_{\alpha\gamma}$
$(\alpha\neq\gamma)$ fields for the corresponding factors. We have
\begin{eqnarray}
&&\exp\Big[{\frac{\beta^2}{4N}\sum_{\alpha}\Big(\sum_i (S^\alpha_i)^2\Big)^2}\Big] 
=\int\prod_\alpha\big(\frac N{4\pi\beta^2}\big)^{\frac 1 2}dq_{\alpha\alpha} 
\nonumber \\
&&\times\exp\Big[ -\frac N{4\beta^2}\sum_\alpha q^2_{\alpha\alpha}
                    +\frac 1 2 \sum_\alpha q_{\alpha\alpha}\sum_i (S_i^\alpha)^2\Big],
\end{eqnarray}
and
\begin{eqnarray}
&&\exp\Big[{\frac{\beta^2}{2N}\sum_{\alpha<\gamma}\Big(\sum_i S^\alpha_i 
                                                            S^\gamma_i\Big)^2}\Big] 
=\int\prod_{\alpha<\gamma}\big(\frac N{2\pi\beta^2}\big)^{\frac 1 2}dq_{\alpha\gamma} 
\nonumber \\
&&\times\exp\Big[ -\frac N{2\beta^2}\sum_{\alpha<\gamma} q^2_{\alpha\gamma}
                    + \sum_{\alpha<\gamma} q_{\alpha\gamma}
		                  \sum_i S_i^\alpha S_i^\gamma\Big].
\end{eqnarray}
We can then integrate over the spin variables to obtain the replicated partition function
as integrals over the HS fields, $q_{\alpha\alpha}$ and $q_{\alpha\gamma}$ and over
the replicated chemical potential $z_\alpha$:
\begin{eqnarray}
&&\overline{Z^n}=
\int\prod_\alpha\big(\frac N{4\pi\beta^2}\big)^{\frac 1 2}
   dq_{\alpha\alpha}
\int\prod_{\alpha<\gamma}\big(\frac N{2\pi\beta^2}\big)^{\frac 1 2}
   dq_{\alpha\gamma} \nonumber \\
&&\times\int_{-i\infty}^{i\infty}
\prod_\alpha \beta\frac{dz_\alpha}{4\pi i}
\exp\Big[-\frac N{4\beta^2}\sum_\alpha q^2_{\alpha\alpha}
         -\frac N{2\beta^2}\sum_{\alpha<\gamma}q^2_{\alpha\gamma}
\nonumber \\
&&+\frac N 2
\Big( \sum_\alpha \beta z_\alpha -
     \mathrm{tr}\ln\big[ (\beta z_\alpha - q_{\alpha\alpha})\delta_{\alpha\gamma}
                  -q_{\alpha\gamma}\big]\Big) \Big],
\end{eqnarray}
where $\mathrm{tr}$ is taken with respect to the replica index.

In the large-$N$ limit, these integrals can be
evaluated by the steepest descent method. For $T>T_c\equiv 1$, the saddle points
are at $q_{\alpha\gamma}=0$, $q_{\alpha\alpha}=\beta^2$ and
$z_\alpha=\beta+\beta^{-1}$. This is the well-known result \cite{KTJ}
from the $N\to\infty$ analysis of the spherical spin glass.
We investigate the finite-size corrections in the limit $T\to T_c=1$
by considering the fluctuations around these saddles.
Writing $q_{\alpha\alpha}=\beta^2+y_\alpha$ and
$\beta z_\alpha=1+\beta^2+i x_\alpha$, we have
\begin{eqnarray}
&&\overline{Z^n_{\mathrm{sp}}}=C\int\prod_{\alpha<\gamma} 
                        \big(\frac N{2\pi\beta^2}\big)^{\frac 1 2} dq_{\alpha\gamma}
                       \int\prod_\alpha \big(\frac N{4\pi\beta^2}\big)^{\frac 1 2}dy_\alpha
		       \nonumber \\
&&\times\int_{-\infty}^\infty\prod_\alpha \frac{dx_\alpha}{4\pi}
\exp\Big[ -\frac N {4\beta^2}
\Big( \sum_{\alpha,\gamma} q^2_{\alpha\gamma} +\sum_\alpha y^2_\alpha \Big)
\nonumber \\
&&\qquad\qquad -\frac N 2 \sum_\alpha (y_\alpha-i x_\alpha)  \\
&&\qquad\qquad -\frac N 2 \mathrm{tr}\ln\big[\{1-(y_\alpha-ix_\alpha)\}
                      \delta_{\alpha\gamma}-q_{\alpha\gamma}\big] \Big],
		      \nonumber
\end{eqnarray}
where $C=\exp((nN/2) (1+\beta^2/2))$. 
If we expand
the logarithm in powers of the fields, we find that the quadratic terms in the diagonal fields 
$x_\alpha$ and $y_\alpha$ inside the exponential are given by
\begin{equation}
-\frac{N}{4}\sum_\alpha
 \Big[(T^2-1)y^2_\alpha +2i x_\alpha y_\alpha +x^2_\alpha\Big].
\end{equation}
Diagonalizing this quadratic form,
we find two eigenvalues with nonvanishing negative real parts at $T=T_c$,
which implies that the diagonal fields are hard modes near $T_c$ and
can be integrated away without encountering divergences.
Therefore the critical behavior is described by
the off-diagonal partition function, which can be written as
\begin{eqnarray}
Z_{\mathrm{off}}&=&\int\prod_{\alpha<\gamma} 
                        \big(\frac N{2\pi\beta^2}\big)^{\frac 1 2} dq_{\alpha\gamma} 
			\exp\Big[-\frac N 4 (T^2-1)\sum_{\alpha,\gamma} q^2_{\alpha\gamma}
\nonumber\\
&&\quad\quad
+\frac N 6 \sum_{\alpha,\beta,\gamma}q_{\alpha\beta}q_{\beta\gamma}q_{\gamma\alpha}
+O(q^4)\Big], 
\label{Zoff}
\end{eqnarray}
where the quartic and higher order terms in $q_{\alpha\gamma}$ all have coefficients
proportional to $N$. A key point of this discussion is to note that,
in the limit where $ t\equiv (T-T_c)/T_c \to 0$ and $N\to\infty$, 
the quartic and higher order terms can be neglected if we keep $N t^3$ finite. 
This can be easily seen by rescaling
$q_{\alpha\gamma}\to q_{\alpha\gamma}/\sqrt{N t}$ in (\ref{Zoff}).
In fact we can show that the off-diagonal partition function 
in this limit is exactly
the same as the zero-dimensional
cubic field theory defined in (\ref{Zn}) and (\ref{seriesZn})
after identifying the expansion parameter as $g^2=1/(N t^3)$ 
and $\tau=T^2-1=t(2+t)\to 2t$ as $t\to 0$. We have
$Z_{\mathrm{off}}(N,\beta)\to Z(\frac 1{N t^3})$ as $N\to \infty$,
$ t\to 0$ and $N t^3\to$ finite. 
A similar observation was made for the critical finite size corrections of
the Sherrington-Kirkpatrick model in
Ref.~\cite{PRS}. (In this paper, the series for the free energy of the toy 
model was given to order $g^4$).

\section{High-order behavior of the expansion coefficients: Toy problem}
\label{analysis}

\subsection{Leading-order behavior}

\begin{figure}
\includegraphics[width=0.4\textwidth]{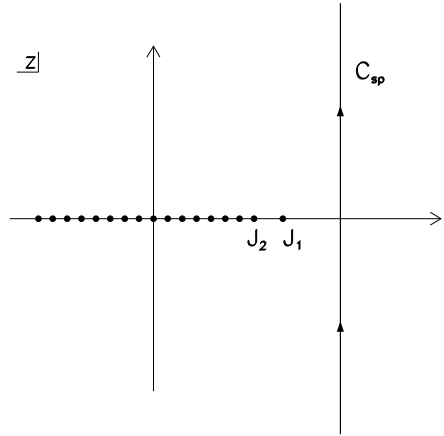}
\caption{The integration contour $C_{\mathrm{sp}}$ used in Eq.~(\ref{Jlambda}).
The filled circles represent the eigenvalues $J_{\lambda}$ schematically. 
The largest and
the second largest eigenvalues are denoted by $J_1$ and $J_2$, respectively.}
\label{Csp}
\end{figure}

Having established the equivalence of the modified spherical spin glass model
to the cubic replica field theory, we can analyze the toy problem (\ref{Zn})
without using the replicas.
Integrating over the spin variables in (\ref{Zsph}), we obtain
\begin{equation}
Z_{\mathrm{sp}}=\beta\int_{-i\infty}^{i\infty} \frac{dz}{4\pi i}
   \exp\Big[\frac{N\beta z} 2
 -\frac 1 2\sum_\lambda\ln\Big( \frac{\beta z} 2 -\frac{\beta J_\lambda}2\Big)
       \Big],
\label{Jlambda}
\end{equation}
where $J_\lambda$ denotes the eigenvalue of the matrix $J_{ij}$.
Note that the contour $C_{\mathrm{sp}}$ of integration lies to the right of the
largest eigenvalue $J_1$. (See Fig.~\ref{Csp}.) 

\begin{figure}
\includegraphics[width=0.4\textwidth]{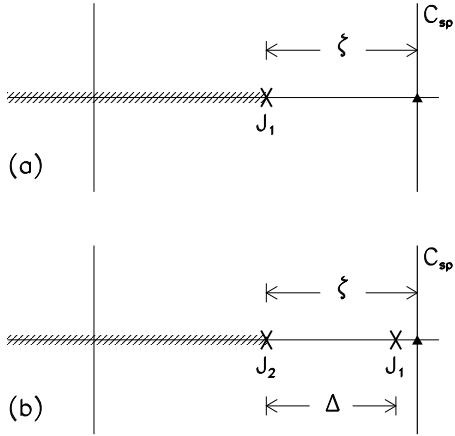}
\caption{(a) The disorder distribution responsible for the dominant
high-order behavior. The largest eigenvalue $J_1$ is far away from the
saddle point that the contour passes through. The rest of the eigenvalues
(the shaded region)
are approximated as a continuum distribution.
(b) The disorder distribution for the sub-dominant high-order behavior.
The separation $\Delta$ between the eigenvalues $J_1$ and $J_2$
is large so that the eigenvalues smaller than $J_2$ can be approximated
as a continuum.}
\label{approx}
\end{figure}

For large $N$, the integral is dominated by the saddle point determined by 
\begin{equation}
\beta=\frac 1 N \sum_\lambda \frac 1 {z-J_\lambda}.
\label{saddle}
\end{equation}
In the limit $N\to\infty$, one can evaluate the sum on the right hand side using
the Wigner semicircle law for the eigenvalue density $\rho(J_\lambda)$ \cite{KTJ}.
Here we are interested in the finite $N$ corrections, in particular, the limit where
$N\to\infty$ and $ t\to 0$ with $N t^3$ held fixed.
In this case, one has to find a solution $z$ of (\ref{saddle})
for given bond realization $J_{ij}$, then calculate the free energy
$-T\ln Z_{\mathrm{sp}}$ from (\ref{Jlambda}), which has to be averaged over the bonds. 
This is an extremely difficult task to carry out analytically.
Instead here we make a self-consistent approximation where we assume the saddle point is
located well away (in the sense to be specified below)
from the largest eigenvalue $J_1$. Since the eigenvalues can be regarded as 
one-dimensional electric charges interacting logarithmically \cite{Dyson,Mehta}, 
an analogy to the 
multipole expansion in electrostatics suggests that 
we can treat the eigenvalues less than $J_1$ as a continuous distribution
given by the semicircle law. That is to take
\begin{equation}
\rho(J_\lambda)\simeq \frac{2(N-1)}{\pi J^2_1}\sqrt{J^2_1-J^2_\lambda}
+\delta(J_\lambda-J_1). 
\end{equation}
In the limit of $N\to\infty$, $ t=(\beta^{-1}-J_1/2)(2/J_1)\to 0$ ($T_c=J_1/2$),
and finite $\eta\equiv N^{1/3} t$, we find that
the distance of the saddle point from $J_1$ is scaled as
$z-J_1\sim O(N^{-2/3})$. In terms of $\zeta\equiv N^{2/3}(z-J_1)(2/J_1)$,
the saddle point equation (\ref{saddle}) reduces to $\eta = \sqrt{\zeta}-1/\zeta$ in this
approximation.
As will be shown below,
the high-order behavior of the perturbation expansion
is described by the small $g^2$ (or large $\eta$) behavior of the partition function.
The distance of the saddle point from the largest eigenvalue
(measured in terms of  $\zeta$) will be large for large $\eta$ and therefore the approximation
 can be
justified for determining the high-order coefficients of the perturbation expansion. 
(See Fig.~\ref{approx} (a).)

Within this approximation, the sum in (\ref{Jlambda}) consists of the term involving $J_1$ and
\begin{eqnarray}
&&\frac 1 {N-1}\sum_\lambda {}^\prime \ln(z-J_\lambda)
 \simeq \frac{z}{J^2_1}\left(z-\sqrt{z^2-J^2_1}\right) \nonumber \\
&&\quad\quad\qquad    +\ln \left[\frac 1 2 \Big(
z+\sqrt{z^2-J^2_1}\Big)\right]-\frac 1 2,
\label{log}
\end{eqnarray}
where the prime indicates the largest eigenvalue is excluded from the sum.
By changing the integration variable to $\zeta$ in (\ref{Jlambda}) and
taking the large-$N$ limit, we obtain
\begin{eqnarray}
Z_{\mathrm{sp}}&\simeq&\frac{\beta e^{-N\beta f_0}}{N^{\frac 1 3}}\;e^{\frac{\eta^3}6}
  \int_{-i\infty}^{i\infty}\frac {d\zeta}{4\pi i}\;
  \frac{\exp(-\frac 1 2 \eta\zeta+\frac 1 3\zeta^{\frac 3 2})}{\sqrt{\zeta}}
  \nonumber \\
 &=&\beta N^{-\frac 1 3} e^{-N\beta f_0}
    \; e^{\frac{\eta^3}6}\int_C
 \frac{d\xi}{2\pi i} \; e^{\frac{\xi^3}3 -\frac{\eta \xi^2}2},
  \label{Z1}
\end{eqnarray}
where $\beta f_0=-\beta J_1 /2 +(1/2)\ln(\beta J_1/4)+1/4+t^3/6$ and
the contour $C$ starts from
$|\xi|=\infty$ with $\arg(\xi)=-\pi /4$ and extends to $|\xi|=\infty$ with
$\arg(\xi)=\pi /4$. We can evaluate the above integral explicitly as
\begin{equation}
Z_{\mathrm{sp}}\simeq
\beta N^{-\frac 1 3} e^{-N\beta f_0}
    \; e^{\frac{\eta^3}{12}} \mathrm{Ai}\big(\frac{\eta^2}4\big) ,
    \label{Airy}
\end{equation}
where $\mathrm{Ai}$ is the Airy function.
We note that the above expression is valid for both $T>T_c$ ($\eta>0$)
and $T<T_c$ ($\eta<0$).
Since the argument of the Airy function is an even function of
$\eta$, we obtain, in the large-$N$ limit,
$\mathrm{Ai}(\eta^2/4)\sim\exp[-|\eta|^3/12]/\sqrt{2\pi|\eta|}$
using the asymptotic behavior of the Airy function.
Therefore, above $T_c$, the leading contribution to the free
energy density, $-N^{-1}\ln Z_{\mathrm{sp}}$,
in the large-$N$ limit is just $\beta f_0$. On the other hand,
below $T_c$, it is given by $\beta f_0 -t^3/6$.
One can explicitly check that these quantities
coincide with the free energy densities given in Ref.~\cite{KTJ}
for the spherical model when $T$ approaches $T_c$ from above and below.

\begin{figure}
\includegraphics[width=0.4\textwidth]{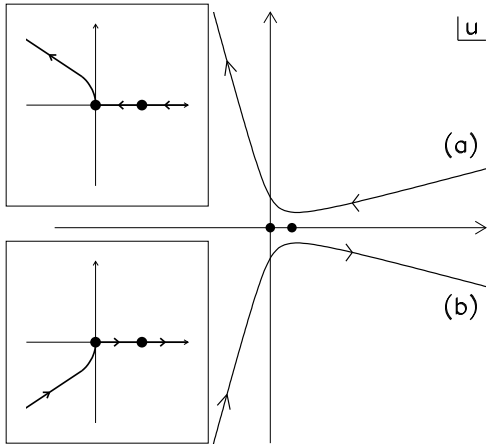}
\caption{The rotated contours used in Eq.~(\ref{Z1u}) for (a) $\arg(g^2)=\pi$
and (b) $\arg(g^2)=-\pi$. The insets show how these contours must be
deformed to pass through the saddle points $u=0$ and $u=1$ (filled circles).}
\label{contouru}
\end{figure}

\begin{figure}
\includegraphics[width=0.4\textwidth]{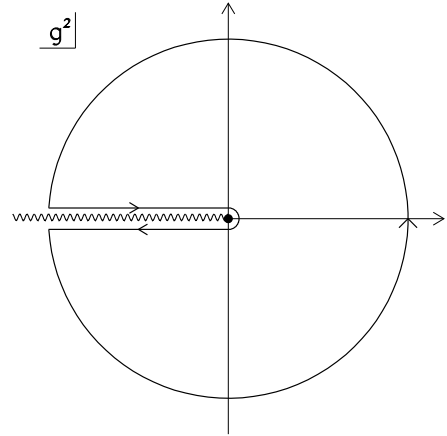}
\caption{The integration contour in the $g^2$ space used to determine the
expansion coefficients $a_K$ in Eq.~(\ref{AK}). The integral reduces to the
one along the branch cut on the $g^2<0$ axis as the contribution from
the circle vanishes when the radius gets large.
}
\label{akg}
\end{figure}

We now consider the finite size corrections
above $T_c$. To do this we introduce $Z_1$ by writing (\ref{Z1}) as
\begin{equation}
Z_{\mathrm{sp}}=\frac{\beta e^{-N\beta f_0}}{\sqrt{2\pi N t}} Z_1.
\label{Z1def}
\end{equation}
Note that the square root in the denominator of (\ref{Z1def})
comes from the Gaussian fluctuations
around the large-$N$ saddle point. The finite size corrections
relevant to the zero-dimensional cubic field theory comes from
$Z_1$, since we can show that
$Z_1=Z_1(g^2)$ is a function
of the expansion parameter $g^2=\eta^{-3}$ only. It is given by
\begin{equation}
Z_1(g^2)=\sqrt{\frac{2\pi}{g^2}}\int_C\frac{du}{2\pi i} \;
\exp\Big[\frac 1{g^2} \Big(\frac{u^3}3 -\frac{u^2}2+\frac 1 6\Big)\Big],
\label{Z1u}
\end{equation}
where $u=g^{2/3} \xi$.
Although $Z_1(g^2)$ can be evaluated analytically as in
(\ref{Airy}), we calculate the high-order
expansion coefficients of $Z_1$ by an indirect method, where the self-consistency
of our approximation is more apparent. (Remember that we are striving to get the behavior of
the high-order coefficients exactly; our approximation of the distribution of
the eigenvalues as a continuum described by the semicircle law does not give the 
correct low-order coefficients in the expansion of the free energy).
From the integral representation (\ref{Z1u}),
we can analytically continue $Z_1(g^2)$ to
any complex $g^2$ by rotating the contour appropriately. 
We find that $Z_1$ has a branch cut along the
$g^2<0$ axis. The imaginary part of $Z_1$ is
discontinuous crossing this axis.
For small $|g|^2$, we can evaluate the
discontinuity by the steepest descent method.
For $\arg(g^2)=\pm\pi$, the contour $C$ is rotated
by $\pm\pi/3$ as shown in Fig.~\ref{contouru}. 
Among the two saddle points $u=0$ and $u=1$ that $C$ can pass through,
the latter produces a real quantity
which is the same for $\arg(g^2)=\pm\pi$, while the former is responsible for
the discontinuous imaginary part
\begin{equation}
\mathrm{Im}Z_1 (g^2;\arg(g^2)=\pm\pi)=\mp\frac 1 2 \exp(\frac 1{6g^2})
\big[1+O(g^2)\big].
\label{ImZ1sp}
\end{equation}
This can be  used  to  extract  the  coefficients  $A_K$  of  the  perturbation
expansion.   We follow  the standard  procedure  \cite{Lipatov,BLZ} by
writing a dispersion  relation for $Z_1(g^2)$ for $g^2>0$  in terms of
an integral  over a contour that  runs around the cut  in the negative
$g^2$  axis. (See Fig.~\ref{akg}.) Therefore,
the  coefficients  $a_K\equiv\lim_{n\to 0} A_K/n$ of the perturbation expansion in
the toy spin glass field theory is given by
\begin{equation}
a_K\simeq\frac 1\pi\int_{-\infty}^0 dg^2\;\frac{\mathrm{Im}
Z_1(g^2;\arg(g^2)=\pi )} {(g^2)^{K+1}}.
\label{AK}
\end{equation}
For large $K$, this integral is dominated by
the saddle point $g^2=-1/(6K)$. This implies that
the information on $\mathrm{Im} Z_1(g^2)$  for small
$g^2$ can be used to obtain $A_K$ for large $K$, which justifies 
the present approximation. We finally obtain
\begin{equation}
a_K \simeq\frac {1} {2\pi}(-6)^K K! K^{-1}\big[1+O(\frac 1 K)\big].
\label{AKn}
\end{equation}
The coefficients with alternating signs are consistent with the
low-order behavior obtained in Sec.~\ref{zero}.

\subsection{Subdominant behavior}

There can be other contributions to the free energy from different disorder 
distributions.  For example, when the eigenvalues are distributed in such a way that 
the saddle point is not 
very far from the largest eigenvalue, the above approximation breaks down. 
In this case, we expect a different behavior of the free energy. To handle this,
we again make an approximation which can be justified self-consistently.
We consider a disorder distribution where the second largest
eigenvalue $J_2$ is well separated from $J_1$ such that the spectrum below
$J_2$ can be described by the semicircle law. (See Fig.~\ref{approx} (b).)
This assumption is justified in the following analysis. We
approximate
\begin{equation}
\rho(J_\lambda)\simeq\frac{2(N-2)}{\pi J^2_2}\sqrt{J^2_2-J^2_\lambda}
+\sum_{i=1,2}\delta(J_\lambda-J_i).
\end{equation}
and
\begin{equation}
Z_{\mathrm{sp}}\simeq\beta\int_{-i\infty}^{i\infty} \frac{dz}{4\pi i}
  \frac{
   \exp\big[\frac{N\beta z} 2
   -\frac N 2 \ln(\frac\beta 2)
 -\frac 1 2\sum_\lambda^{\prime\prime}\ln(z-J_\lambda)
       \big]}
 {\sqrt{z-J_1}\sqrt{z-J_2}},
\label{J1J2}
\end{equation}
where the double-primed sum excludes $\lambda=1$ and 2.
This sum can be evaluated as in (\ref{log}) with $J_2$ replacing $J_1$.
We make the same series of integration variable changes
leading to (\ref{Z1}) and (\ref{Z1u}) (using $\zeta\equiv N^{2/3}(z-J_2)(2/J_2)$
and $T_c=J_2/2$ in this case), and take the large-$N$ limit. We obtain
$Z_{\mathrm{sp}}\simeq\beta e^{-N\beta f_0} Z_2$ where $f_0$ is the same as before
with $J_1$ replaced by $J_2$ and
\begin{equation}
Z_2(g^2,\Delta)=\int_C
 \frac{du}{2\pi i} \; \frac{\exp[\frac 1 {g^2}
 (\frac{u^3}3 -\frac{ u^2}2+\frac 1 6)]}
 {\sqrt{u^2-g^{\frac 4 3}\Delta}},
  \label{Z2}
\end{equation}
with the eigenvalue spacing $\Delta\equiv N^{2/3}(J_1-J_2)(2/J_2)$. 
(Recall $g^{-2}=\eta^3$.)

The contribution from this arrangement of disorder to the free energy,
which we denote by $F_{\mathrm{sub}}$,
is obtained by averaging $-\ln Z_2$ over
the distribution $p(\Delta)$ of the eigenvalue spacing $\Delta$. Among the
subdominant contributions to the high-order behavior, we focus on those from
possible zeros of $Z_2$. By explicitly
evaluating the contour integral (\ref{Z2}) numerically
for given $g$, we find that there exists
a complex conjugate pair of zeros, $\Delta_0$ and $\Delta^*_0$ 
in the complex-$\Delta$ plane.
To make analytic progress on the contribution from these zeros,
we look at the fluctuation around the saddle
point $u=1$. By writing $u=1+ig y$ and neglecting $O(g)$ terms, we obtain
$Z_2 \simeq (\sqrt{g/2})\widetilde{Z}_2(g^2,\Delta)$, where
\begin{equation}
\widetilde{Z}_2(g^2,\Delta)=\widetilde{Z}_2(v)=
\int_{-\infty}^\infty \frac{dy}{2\pi}\;
\frac{e^{-y^2/2}} {\sqrt{v+iy}}
\label{tildeZ2}
\end{equation}
with $v=(1-g^{4/3}\Delta)/(2g)$. Note that $v$ is assumed to
be of $O(1)$, which means $\Delta\sim O(g^{-4/3})$.
This is consistent with the present approximation where
the separation of eigenvalues $\Delta$ is very large. Writing
\[
\frac 1{\sqrt{v+iy}}=\int_{-\infty}^{\infty} \frac{ds}{\sqrt{2\pi}}\;
e^{-s^2(v+iy)}
\]
and integrating over $y$ in (\ref{tildeZ2}), we have
\begin{equation}
\widetilde{Z}_2=\int_{-\infty}^\infty \frac{ds}{2\pi}\;
e^{-vs^2-s^4/2}=\sqrt{\frac{v}{2}}\frac{e^{v^2/4}}{2\pi}
K_{\frac 1 4}(\frac{v^2}4).
\end{equation}
The zeros of $\widetilde{Z}_2$ come from the infinitely many
zeros of the modified Bessel function
$K_{\frac 1 4}$ \cite{McKane94,Alvarez}. We can arrange them as complex
conjugate pairs, $v_m$ and $v^*_m$, $m=0,1,2,\ldots$, given approximately
for large $m$ by
\begin{equation}
v^2_m\sim e^{3\pi i/2}\big[-i\ln 2+(4m+3)\pi\big]
\end{equation}
and for $m=0$ exactly (up to four decimal places)
by $v^2_0=9.4244-0.6928 i$.
Among the infinitely many zeros, we focus only on the pair $v_0$ and $v^*_0$
closest to the origin as our numerical evaluation of zeros of
(\ref{Z2}) is consistent with $\Delta_0=(1-2gv_0)/g^{4/3}$ for small $g$.
The other zeros probably give only subdominant contributions compared to the
first ones.

\begin{figure}
\includegraphics[width=0.4\textwidth]{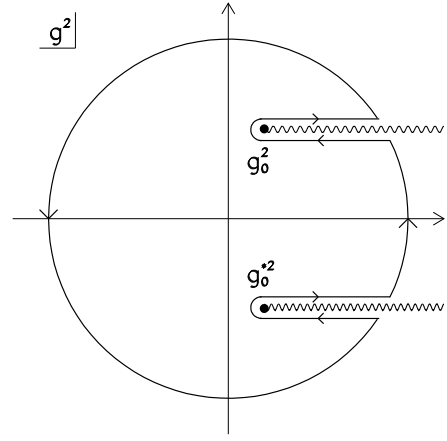}
\caption{The deformed contour in $g^2$ space that leads to Eq.~(\ref{asub}).
The branch points $g_0^2$ and $g_0^{*2}$ correspond to
the zeros of the logarithm.}
\label{contourg}
\end{figure}

The subdominant contribution to the expansion coefficient denoted by
$\widetilde{a}_K$ can be calculated as
\begin{equation}
\widetilde{a}_K=\frac 1{2\pi i}\oint\frac {d g^2}{(g^2)^{K+1}}\;F_{\mathrm{sub}},
\end{equation}
where the integral is over a closed contour surrounding the origin.
By interchanging the order of integration, we can write
\begin{equation}
\widetilde{a}_K=-\frac 1{2\pi i}\int_{0}^{\infty} d\Delta \;p(\Delta)
\oint\frac {d g^2}{(g^2)^{K+1}}\ln \widetilde{Z}_2(v).
\end{equation}
The integral over $g^2$ is done by deforming the contour such that it runs
along a circle of radius $R$ and
around the branch cuts associated with the zeros, $g^2_0$ and
$g^{*2}_0$ of $\widetilde{Z}_2$ for fixed $\Delta$. (See Fig.~\ref{contourg}.)
The integral along the circle vanishes as $R\to\infty$ and we have
\begin{eqnarray}
\widetilde{a}_K&=&\int_{0}^{\infty} d\Delta \;p(\Delta)
\Big[\int_{g^2_0}^\infty\frac {d g^2}{(g^2)^{K+1}}
     +\int_{g^{*2}_0}^\infty\frac {d g^2}{(g^2)^{K+1}}\Big] \nonumber\\
                     &=&\frac 1 K \int_{0}^{\infty} d\Delta \;p(\Delta)
\Big[\frac 1 {(g^2_0)^K}+\frac 1 {(g^{*2}_0)^K}\Big].
\label{asub}
\end{eqnarray}
The eigenvalue spacing distribution is known \cite{Mehta} to take the form
$p(\Delta)\sim\exp(-(\pi^2/16)(\Delta/D)^2))$ for large $\Delta$
when the mean eigenvalue spacing is $D$.
From the numerical results in Ref.~\cite{RM} on the eigenvalue
spacing distribution of real symmetric matrices, we can calculate
the mean spacing near the edge of the spectrum as $D\simeq 2.30$.
The integral over $\Delta$ in (\ref{asub}) can be done using the steepest descent method
for large $K$. The saddle point for the first term in (\ref{asub}) is given by
\begin{equation}
-\frac{\pi^2}{8D^2}\Delta-\frac{2K}{g_0}\Big(\frac{dg_0}{d\Delta}\Big)=0,
\label{s1}
\end{equation}
where $d g_0/d\Delta$ can be obtained from the defining equation
\begin{equation}
1-g^{\frac 4 3}_0\Delta=2g_0 v_0 .
\label{s2}
\end{equation}
We can solve (\ref{s1}) and (\ref{s2}) for large $K$ to obtain the saddle point
values
\begin{eqnarray*}
&&g^{\mathrm{sad}}_0=\alpha^{\frac 3 8}K^{-\frac 3 8}
\Big[ 1-\Big( \frac{5 v_0}{4\alpha^{\frac 5 8}}\Big)  K^{-\frac 3 8}
+O(K^{-\frac 3 4})\Big], \\
&&\Delta^{\mathrm{sad}}=\alpha^{-\frac 1 2}K^{\frac 1 2}
\Big[1-2\big(1-\frac 5 {6\alpha}\big)v_0\alpha^{\frac 3 8}K^{-\frac 3 8}
+O(K^{-\frac 3 4})\Big],
\end{eqnarray*}
where $\alpha=\pi^2/(12D^2)\simeq 0.155$. Note that the large-$K$ behavior corresponds to
small $g$ and
large $\Delta$ with $\Delta\sim g^{-4/3}$, which means that these results are
entirely in the regime of the present approximation.
Inserting these into (\ref{asub}) we finally obtain
\begin{eqnarray}
\tilde{a}_K&\sim &\left( K! \right)^{\frac 3 4} a^K \exp\left[
                   -bK^{\frac 5 8}+O(K^{\frac 1 4})\right] \nonumber \\
                 &&\qquad\times\cos\left(cK^{\frac 5 8}+O(K^{\frac 1 4})\right),
		 \label{aksub}
\end{eqnarray}
where  $a=\alpha^{-3/4}\simeq 4.05$, $b=-3\alpha^{3/8}\mathrm{Re}(v_0)
\simeq   3.15$   and   $c=3\alpha^{3/8}\mathrm{Im}(v_0)\simeq   3.38$.
Compared  with   (\ref{AKn}),  this  is   a  subdominant  contribution
containing only  a fractional power  of $K!$. The coefficients  do not
alternate  in sign  as  in  (\ref{AKn}) but  oscillate  with a  cosine
function with  an increasing periodicity. The situation  is similar to
that     in    the     zero-dimensional     disordered    ferromagnets
\cite{BMMRY,McKane94,Alvarez},  where this  type  of oscillation  also
occurs in  the subdominant  terms. We  expect that as  in the  case of
disordered ferromagnets the subdominant  terms make the resummation of
the series non-trivial such  that a straightforward Borel summation is
spoiled. However, it seems likely that the series could be resummed in
other ways and anyway,  the evidence from Ref.~\cite{BMMRY} suggests
that the  straightforward Pad\'{e}-Borel  method works well  for short
series even in the presence of subdominant terms.

There are obviously  other types of bond distribution
which could give rise to subdominant contributions to the high-order behavior of
the perturbation series besides the one studied in this subsection. We suspect that
the type studied here provides the largest of these contributions but we have
no proof of this.

\section{Replica Approach to High Order Behavior in the Toy Problem}
\label{ToyReplica}

While  the  mapping  of  the  toy  integral,  Eq.~(\ref{Zn}),  to  the
spherical  model has  enabled us  to obtain  direct estimates  for the
high-order  terms of its  perturbative expansion,  in order  to obtain
high-order estimates  for the  $d$-dimensional field theory  and hence
for  critical exponents we  have to  discover how  to obtain  the same
high-order  estimates  directly  from  the  integral  in  the  replica
variables $q_{\alpha\beta}$. Once this  has been done the extension to
field theory is  relatively straightforward and is carried  out in the
next  Section. Unfortunately  the direct  replica approach  is neither
obvious nor rigorous. Without the results obtained from the mapping to
the spherical  model we  would have had  no confidence in  the replica
procedure which we were forced to use.

It is useful to first examine the integral for
the special case of $n=3$ when the integrals can be done explicitly
and exist if $w$ is pure imaginary. This case was analyzed in
Ref.~\cite{McKane}.
Here we follow the same analysis to clarify some points
which will be important in the case of general $n$.
For $n=3$, we have after setting $\tau=1$ for simplicity
\begin{equation}
Z_3(w)=\frac{4\pi}{(2\pi)^{3/2}}\int_0^\infty dR R^2
e^{-\frac{R^2}2}f(wR^3),
\label{Z3}
\end{equation}
where
\begin{eqnarray}
f(wR^3)&=&\frac 1{4\pi}\int_0^{2\pi}d\phi\int_0^\pi d\theta\sin\theta
e^{wR^3\sin^2\theta\cos\theta\cos\phi\sin\phi} \nonumber\\
&=&\frac\pi 3 I_{\frac 1 6}\left(\frac{wR^3}{6\sqrt{3}}\right)
I_{-\frac {1} 6}\left(\frac{wR^3}{6\sqrt{3}}\right)
\end{eqnarray}
with the modified Bessel function $I_\nu(x)$.
For pure imaginary coupling $w=iw^\prime$,
$w^\prime$ real, $f=(\pi/3)J_{1/6}(w^\prime R^3/6\sqrt{3})
J_{-1/6}(w^\prime R^3/6\sqrt{3})$ with the ordinary Bessel function
$J_\nu(x)$. We can easily see that
the integral in (\ref{Z3})
is well-defined and real in this case.

When analytically continued to real coupling $w$, $Z_3$ develops
an imaginary part which is discontinuous crossing
the branch cut along the real $w$ axis. For small values of coupling,
one can calculate the discontinuity
using the steepest descent method
on the integral in (\ref{Z3}). Since the main contribution
to the imaginary part comes from the saddle point $R\sim \mathrm {O}(1/w)$,
we first study the asymptotic behavior of $f$ when $|wR^3|$ is very large, 
which is
\begin{equation}
f(wR^3)\sim\frac{\sqrt{3}}{wR^3}\left[\exp\left(\frac{wR^3}{3\sqrt{3}}\right)
-\exp\left(-\frac{wR^3}{3\sqrt{3}}\right)+\sqrt{3}i\right].
\label{asympt}
\end{equation}
The above  expansion is valid  for $0\leq \arg  (w) \leq \pi$,  or for
$0\leq  \arg (w^2)  \leq  2\pi$.   For $\arg  (w)=0$,  only the  first
exponential  in  (\ref{asympt})  is  important  and  the  integral  in
(\ref{Z3})   is   dominated  by   the   saddle   points  $R_1=0$   and
$R_2=\sqrt{3}/w$.   The  steepest  descent   direction  at   $R_2$  is
perpendicular to the real-$R$  axis along which the integral generates
the imaginary part.  We can deform the contour in (\ref{Z3}) such that
it starts from $R_1$ along the positive real-$R$ axis toward $R_2$ and
makes an upward  turn at $R_2$. (Examples of  similar deformations are
in Ref.~\cite{McKane3}.)  We note that,  since we only pass a half of the
steepest  descent path  of $R_2$  in this  way, the  Gaussian integral
coming from the fluctuation around  $R_2$ produces a half of the total
fluctuation  contribution.   Keeping this  in  mind,  we evaluate  the
integral to obtain $\mathrm{Im}Z_3(w)=\exp(-\frac 1 {2w^2})$ for $\arg
(w)=0$.  For $\arg (w)=\pi$,  the saddle points are $R^\prime_1=0$ and
$R^\prime_2=-\sqrt{3}/w$ and the continuation  of the contour used for
$\arg (w)=0$ to this case is  the one which makes the downward turn at
$R^\prime_2$. We finally obtain the discontinuity in the imaginary part
along the branch cut on the real-$w$ axis as
\begin{equation}
\mathrm{Im}Z_3(w)=\pm\exp(-\frac 1 {2w^2}),
\label{ImZ3}
\end{equation}
where the positive and negative signs
correspond to $\arg (w^2)=0$ and $\arg (w^2)=2\pi$, respectively.
This exponentially small imaginary contribution to the partition function
at the physical coupling can
be used to determine the large order behavior of the perturbation
expansion, but at the same time its presence indicates that the cubic field theory 
for $n=3$ is ill-defined and requires stabilizing quartic terms for its existence.

We now study the case of general $n$. For small values of
the expansion parameter, $w^2/(\tau/2)^3$, we can evaluate
the integrals in Eq.~(\ref{Zn}) using the steepest descent
method. Saddle points are found by solving
\begin{equation}
-\tau q_{\alpha\beta}+w\sum_{\gamma}q_{\alpha\gamma}q_{\gamma\beta}=0.
\label{fsaddle}
\end{equation}
The trivial solution of this equation $q_{\alpha\beta}=0$ is the starting
point of the perturbative expansion. Non-perturbative terms arise from its
non-trivial solution. The set of saddles which we study are
$q_{\alpha\beta}=q$,
when both $\alpha$ and $\beta$ lie in the interval between $1$ and $r$,
where  and $q=\tau/(r-2)w$. This can be described
schematically as
\begin{eqnarray}
q_{\alpha\beta}&=&\left(
\begin{array}{ccc|cc}\cline{1-3}
\multicolumn{1}{|c}{}&&&& \\
\multicolumn{1}{|c}{}&\multicolumn{1}{c}{\;q\;\;}&&\multicolumn{1}{c}{\;\;0}&
\\
\multicolumn{1}{|c}{}&&&& \\ \hline
&\multicolumn{1}{c}{0}&&\multicolumn{1}{c}{\;\;0}&
\end{array}
\right). \\
&&\quad\underbrace{\quad\quad\quad}_r
\underbrace{}_{n-r} \nonumber
\end{eqnarray}
Clearly other types of saddle exist besides this.
We first focus on this class of saddles, which we call scheme I.
The Hamiltonian density for this saddle is given by
\begin{equation}
\mathcal{H}_s=\frac{r(r-1)\tau^3}{12(r-2)^2 w^2}.
\label{hs}
\end{equation}
There are more solutions to
Eq.~(\ref{fsaddle}) with the same Hamiltonian (\ref{hs}), which can be
obtained by switching the signs of some of $q_{\alpha\beta}$.
One can take, for example, $q_{1\beta}=-q$ for $\beta\neq 1$, and
$q_{\alpha\beta}=q$ for $\alpha,\beta\neq 1$. Since one can pick
any other subscript than 1, the number of such solutions is $r$.
One can also switch the signs of two
different sets such as $q_{1\gamma}=q_{2\gamma}=-q$ for $\gamma\neq 1,2$
and keep all the other elements $q_{\alpha\beta}=q$.
The number of such solutions is $r(r-1)/2$.
Similarly, one can switch the signs of $3,4,\ldots, r-1$
different sets of $q_{\alpha\beta}$. However, we can show that
the solutions obtained by switching the signs of $k$ different sets
are equivalent to those from picking $r-k$ different sets for
$k=1,2,\ldots,r-1$.
Therefore, the total number $S(r)$ of the solutions with
the Hamiltonian (\ref{hs}) is
\begin{equation}
S(r)=\sum_{k=0}^{[r/2]}{}^r C_k=\left\{
\begin{array}{cc}
2^{r-1},& r\quad \textrm{odd}, \\
2^{r-1}+\frac{r!}{2\left[\left( r /2\right)!\right]^2}, & r\;\;
\textrm{even},
\end{array}
\right.
\end{equation}
where $[x]$ denotes the largest integer less than or equal to $x$.

We need to include the Gaussian fluctuations around these saddles. The
matrix of the second derivatives has six distinct eigenvalues: a
``breather'' mode with eigenvalue $-\tau$ which is non-degenerate, one other
negative eigenvalue, $-\tau/(r-2)$ which is $(n-r)$-fold degenerate, and
eigenvalues $2\tau/(r-2)$, $(r-1)$-fold degenerate, $r\tau/(r-2)$,
$(r(r-3)/2)$-fold degenerate,
$(2r-3)\tau/(r-2)$, $(r-1)(n-r)$-fold degenerate, and $\tau$,
$(n-r)(n-r-1)/2$-fold degenerate.
Setting $\tau=1$ for simplicity and collecting the contributions
from the saddle points to Gaussian order, we obtain
\begin{eqnarray}
&&Z^{(\mathrm{I})}(w^2)=[1+\mathrm{O}(w^2)]  \label{Znsaddle}\\
&&+\frac 1 2 \sum_{r=3}^n {}^nC_r \; S(r)
\exp\left[-\frac{r(r-1)}{12(r-2)^2w^2}\right](-1)^{\frac 1 2}
\nonumber\\
&&\times
\left(\frac{r-2}{2}\right)^{\frac{r-1}2}
\left(\frac{r-2}{r}\right)^{\frac{r(r-3)}4}
\big(-(r-2)\big)^{\frac{n-r}2}\nonumber \\
&&\times\left(\frac{r-2}
{2r-3}\right)^{\frac{(n-r)(r-1)}2}\big(1\big)^{\frac{(n-r)(n-r-1)}4}
[1+\mathrm{O}(w^2)], \nonumber
\end{eqnarray}
where the first term corresponds to the usual perturbation expansion
in $w$. The sum over $r$ starts at  3 since in Eq.~(\ref{fsaddle}) the index $\gamma$ must
differ from both $\alpha$ and $\beta$. The factor ${}^nC_r$ denotes
the number of ways of introducing $r$ non-zero blocks.
Collecting the contributions from the saddles in this way, instead of
deforming a contour in a multi-dimensional complex space, determines the
partition function up to an overall factor.
The analysis of the $n=3$ case suggests that there is
an overall factor of $1/2$ coming from the fact that, for
the nontrivial saddles, only a half of the Gaussian integrals contribute
compared to the perturbative one.
Indeed, for $n=3$, one can explicitly check that
the non-perturbative part of (\ref{Znsaddle}) reduces to Eq.~(\ref{ImZ3}).
The negative eigenvalues are responsible for the factor $(-1)^{(n-r+1)/2}$,
which generates the imaginary part in $Z$.

For finite $n$, the saddles in the scheme I correspond to the
partition function which is well defined except for real $w$ resulting in
a branch cut on the positive $w^2$ axis. The discontinuity of the imaginary
part across the cut is exponentially small
$\sim\exp(-n(n-1)/(12(n-2)^2w^2))$. (The imaginary part of
(\ref{Znsaddle}) is dominated by the $r=n$ term,
since the Hamiltonian (\ref{hs}) decreases monotonically as
$r$ increases.) If the cubic spin glass field theory is well-defined for
real coupling $w$, we expect that the cut moves to the negative $w^2$ axis
as we take the $n\to 0$ limit and that there is an exponentially small
discontinuity across the negative $w^2$ axis. The migration of the cut as
the analytic continuation of $n\to 0$ is taken is exactly what happens
in the percolation problem \cite{McKane}. In that case, the Hamiltonian
$\mathcal{H}_s$ for
the saddles depends explicitly on $n$ such that it changes
sign as $n\to 0$. In the present case, $\mathcal{H}_s$ in (\ref{hs}) is
independent of $n$, and there is no way of producing an exponentially 
small discontinuity across the negative $w^2$ axis from these saddles.
Therefore, we conclude that the saddles in the scheme I
are not sufficient to describe the partition function
in the $n\to 0$ limit.

This observation leads us to consider
another type of solution of Eq.~(\ref{fsaddle}), which we call
scheme II. It is inspired by the replica symmetry
breaking scheme used by two of us \cite{AM} to describe the free
energy fluctuations. Instead of taking all
$q_{\alpha\beta}$ nonzero for $\alpha,\beta=1,2\ldots r$ as in
the previous scheme,
we set $q_{\alpha\beta}=q$ only when $\alpha$ and $\beta$ belong to the $m$
blocks of size $r/m$ on the diagonal of the $r\times r$ matrix as
\begin{eqnarray}
q_{\alpha\beta}&=&\left(
\begin{array}{cccc|cc}\cline{1-1}
\multicolumn{1}{|c|}{\;q}&&&&& \\ \cline{1-2}
&\multicolumn{1}{|c|}{\;q}&&&& \\ \cline{2-2}
&&\ddots &&& \\ \cline{4-4}
&&&\multicolumn{1}{|c|}{\;q} && \\ \hline
&&&&~~&~~ \\
&&&&~~&~~
\end{array}
\right). \\
&&\;\;\;\underbrace{\quad\quad\quad\quad\quad\;\;\;}_r\:
\underbrace{\quad\quad\;}_{n-r} \nonumber
\end{eqnarray}

The key point of the construction of these saddles is that
we let $m\to\infty$ before $n\to 0$. For finite $m$, this scheme
is just a generalization of the scheme I which produces only subleading terms.
For $m\to\infty$, however,
the Hamiltonian becomes
\begin{equation}
\mathcal{H}_s=\frac{m(\frac r m) (\frac r m -1)\tau^3}{12(\frac r
m-2)^2 w^2}
\to
-\frac{r\tau^3}{48 w^2},
\label{hs2}
\end{equation}
which has the opposite  sign to the one in the scheme  I. This can now
describe the  partition function where  the cut lies on  the imaginary
$w$     axis.       The     solution     to      (\ref{fsaddle})     is
$q=\tau/(r/m-2)w\to-\tau/2w$  as  $m\to\infty$.  

The  matrix  of  the
second  derivatives  necessary to  include  the Gaussian  fluctuations
around the  saddles, has  two distinct positive  eigenvalues, $\tau/2$
and    $\tau$,   which    are,    respectively,   $r(n-r)$-fold    and
$(n-r)(n-r-1)/2$-fold  degenerate.   There   exist  a  null  eigenvalue,
$r(r-3)/2$-fold  degenerate,  and  one  negative  eigenvalue  $-\tau$,
$r$-fold degenerate.  This negative  eigenvalue is responsible for the
factor $(-\tau)^{-\frac  r 2}$ which  generates the imaginary  part for
odd values of $r$.  The total number of solutions that can be obtained
by switching the sign  of $q_{\alpha\beta}$ is $[S(r/m)]^m\to 2^{r/2}$
as $m\to  \infty$.  (This  limit exists if  we assume $r/m$  is even.)
From  (\ref{hs2}), we  can see  that  the saddles  with smallest  $r$
dominate for pure imaginary  $w$.  Therefore, the leading contribution
to the imaginary  part of (\ref{Zn2}) should come from  
the first i.e. $r=1$
term as we can see no reason why it should be excluded in this kind of
replica symmetry breaking scheme. Thus
\begin{eqnarray}
\mathrm{Im}    Z^{\rm    (II)}(w^2)&=&    \pm\frac    {{}^nC_1}    2
\exp\left[\frac {\tau^3}{48w^2}\right] \; 2^{\frac 1 2} \;\left(\frac
2  \tau\right)^{\frac 1  2 (n-1)}  \nonumber\\ &\times&  \left(\frac 1
\tau \right)^{\frac 1 4 (n-1)(n-2)+\frac 1 2} [1+O(w^2)], \label{Zn2}
\end{eqnarray}
where  the upper and  lower signs  correspond to  $\arg(w)=-\pi/2$ and
$\arg(w)=\pi/2$, respectively.

We  take   the  $n\to   0$  limit  of   (\ref{Zn2})  using   the  fact
${}^nC_r=n(-1)^{r-1}/r+O(n^2)$,    which   can    be    derived   from
$1/\Gamma(n-r+1)=\Gamma(r-n)\sin(\pi(r-n))/\pi$.   We  finally  obtain
the  discontinuity of  the imaginary  part of  the  partition function
(\ref{Zn}) in the  limit $n\to 0$ across the cut  on the imaginary $w$
($w^2<0$) axis as
\begin{equation}
\lim_{n\to  0}  \mathrm{Im}\frac{Z^{(\mathrm{II})}(w^2)}n  \sim \pm
\exp\left(\frac {\tau^3}{48w^2}\right).
\label{ImZ2}
\end{equation}
We cannot obtain the precise prefactor to the 
exponential term because we have neglected the contributions of the 
massless eigenvalue. In principle, this soft mode could be integrated out
by identifying the underlying symmetries associated with the saddle 
point. This does not seem obvious to us.
 However, its contribution is subdominant to that from the exponential and we shall ignore 
its contribution.

The discontinuity in the imaginary  part of the partition function can
be  used  to  extract  the  coefficients  $A_K$  of  the  perturbation
expansion. To do that we write the above result
in terms of the expansion parameter $g^2$. Recalling
$\tau=2t$, we can write the right hand side of Eq.~(\ref{ImZ2})
as $\pm\exp(1/6g^2)$ for $\arg(g^2)=\mp\pi$. This is the same as
Eq.~(\ref{ImZ1sp}) up to the undetermined prefactor.
Therefore we obtain exactly the same high-order behavior as in (\ref{AK})
\begin{equation}
\lim_{n\to 0}\frac {A_K}n \sim (-6)^K K!.
\label{AKn2}
\end{equation}
Without the spherical model mapping one would have had reservations about
the likely correctness of the replica procedure
used.

\section{High order terms of the $\epsilon$ expansion}
\label{epsilon}

The starting point for obtaining the large order form of the $\epsilon$ expansion is the Hamiltonian
of Eq.~(\ref{FT}) but without the quartic terms:
\begin{align}
\mathcal{H}&=\frac 14\sum_{\alpha,\beta} (\nabla q_{\alpha\beta})^2
           +\frac \tau 4\sum_{\alpha,\beta}q_{\alpha\beta}^2
           -\frac w6
\sum_{\alpha,\beta,\gamma}q_{\alpha\beta}q_{\beta\gamma}q_{\gamma\alpha}.
\label{FTD}
\end{align}

Our treatment closely follows that of McKane \cite{McKane}. The saddle points which are the analogue
of Eq. (\ref{fsaddle}) of the toy replica calculation are the instantons which satisfy the equation
\begin{equation}
\nabla^2q_{\alpha\beta} =-\tau q_{\alpha\beta}+w\sum_{\gamma}q_{\alpha\gamma}q_{\gamma\beta}.
\label{instanton}
\end{equation}
For evaluating the high-order coefficients in the $\epsilon$ expansion we can set $\tau=0$ and
look for a solution of the form $q_{\alpha\beta}=w^{-1}d_{\alpha\beta}\phi_c(r)$ in $d=6$
dimensions. Such a solution,
which decouples replica indices from the spatial dependence $r$, exists if
\begin{equation}
d_{\alpha\beta}=\sum_{\gamma}d_{\alpha\gamma}d_{\gamma\beta},
\label{indices}
\end{equation}
and
\begin{equation}
\nabla^2\phi_c(\mathbf{r})=\phi^2_c(\mathbf{r}).
\label{spatial}
\end{equation}
There are spherically symmetric solutions of Eq.~(\ref{spatial}):
\begin{equation}
\phi_c(r)=-\frac{24\lambda^2}{[\lambda^2r^2+1]^2},
\label{spatialsoln}
\end{equation}
where the parameter $\lambda$ reflects the dilatation invariance of Eq.~(\ref{spatial}). We shall
take for $d_{\alpha\beta}$ the replica symmetry broken solution of scheme II. Then the energy of
the instanton is
\begin{eqnarray}
E&=&\int d^6\mathbf{r}\; \Big[
\frac 1 4 \sum_{\alpha,\beta} \big(\nabla \phi_c(r)\big)^2 d_{\alpha\beta}^2/w^2
\nonumber \\
           &&\qquad -\frac w 6
\sum_{\alpha,\beta,\gamma}d_{\alpha\beta}d_{\beta\gamma}d_{\gamma\alpha}
\phi^3_c(r)/w^3\Big].
\label{instantenergy}
\end{eqnarray}
Using the result that
\begin{equation}
\int d^6\mathbf{r} \;\big(\nabla \phi_c(r)\big)^2
=-\int d^6\mathbf{r} \;\phi^3_c(r)=\frac{1152\pi^3}5,
\label{integrals}
\end{equation}
the energy of the instanton is
\begin{equation}
E=\left(\frac{1152\pi^3}{5}\right)\frac{1}{48w^2}=\frac{3}{40g_R^2}
\label{result}
\end{equation}
where $g_R^2=K_6w^2$ and  $K_6=S_6/(2\pi)^6$ and $S_6=\pi^3$ is the surface area of a
six dimensional sphere of unit radius. 

The leading terms in the large order behavior of the $\epsilon$ expansion are obtained by replacing
$g_R^2$ by its fixed point value, which to lowest order is $\epsilon/2$ \cite{Green} and by the
usual saddle-point arguments the coefficient of $\epsilon^K$ for large $K$ for
 any critical exponent goes like
\begin{equation}
\sim K!\left(-\frac{20}{3}\right)^K
\label{highK}
\end{equation}

The next most dominant term is a factor of the form $K^b$. The value of $b$ depends on the critical
exponent being studied and is beyond the scope of this paper. To determine its value
 a treatment is needed of the massless modes which arise in the Gaussian fluctuations around the
 instanton solution.

Inspection of the first three  terms in the $\epsilon$ expansion for $\eta$ and 
$\nu^{-1}-2+\eta$ show that these terms are not growing anything like as rapidly as
 predicted at large $K$. It is perhaps not surprising, therefore, that a
 Pad\'{e}-Borel analysis of the series does not yield good numerical values for the critical 
exponents in three dimensions.

\section{Discussion}
\label{conclusion}

In summary we have studied the nature of the perturbation expansion of the zero-dimensional
cubic replica field theory of spin glasses. By mapping this to the problem of critical
finite-size corrections in a modified spherical spin glasses, we have determined the high-order
behavior of the perturbation expansion coefficients. To the leading order, the coefficients
alternate in sign, but there is a subleading contribution where the terms in the perturbation
series show a cosine-like oscillation. In practice, the effects of these sub-dominant terms
will  be small, making a simple Pad\'{e}-Borel resummation of the series useful, as was 
found to be the case in a similar situation for the disordered ferromagnet \cite{BMMRY}.

Non-perturbative terms are also present in spin glasses. These are 
Griffiths singularities and arise from regions where the values of the
couplings $J_{ij}$ produce a smaller amount of frustration and hence a
locally enhanced  transition temperature.  A discussion of  their form
has  been given in  Ref.~\cite{gs}.  Similar singularities  exist for
disordered ferromagnets  and it is widely believed  that their effects
are very small.  To our knowledge no quantitative  discussion of these
singularities has been  made for spin glasses and  their study remains
to be done. (The toy  problem, because it is zero-dimensional, is free
of Griffiths singularities).

The   $\epsilon$   expansion   for   the  critical   exponents   gives
disappointing  results as regards applications to real spin glasses.
  This  is  not  just  due  to  the  fact  that
$\epsilon=3$ in three dimensions  as in Ref.~\cite{AKM} good results
were obtained  for the exponents  of the percolation problem  in three
dimensions  from  an $\epsilon$  expansion  with  the  same number  of
terms. We do not understand the origin of this problem.

However, to our mind, the most significant remaining problem is what motivated 
this entire study. Namely, does perturbation theory (i.e. the loop expansion)
work well in the spin glass phase or does the existence of ``droplets'' in finite
dimensional spin glasses indicate that it fails completely? Our work does indicate
though that perturbation theory is useful in the high-temperature phase.

\begin{acknowledgements}
We would like to thank Dr Alan McKane for his help and advice on all aspects of
this work. TA acknowledges financial
 support of the European Community's Human Potential Programme under contract 
HPRN-CT-2002-00307, DYGLAGEMEN.
\end{acknowledgements}


\end{document}